\documentclass
[twocolumn,showpacs,preprintnumbers,amsmath,amssymb]
{revtex4}
\usepackage{graphicx}
\usepackage{dcolumn}
\usepackage{bm}
\usepackage{here}
\usepackage{color}
\usepackage{braket}
\usepackage{amsmath}  
\usepackage{amssymb}  
\usepackage{fancybox}

\begin{document}

\newcommand{\vc}[1]{\mbox{\boldmath $#1$}}
\newcommand{\fracd}[2]{\frac{\displaystyle #1}{\displaystyle #2}}
\newcommand{\red}[1]{\textcolor{red}{#1}}
\newcommand{\blue}[1]{\textcolor{blue}{#1}}
\newcommand{\green}[1]{\textcolor{green}{#1}}



\def\ni{\noindent}
\def\nn{\nonumber}
\def\bH{\begin{Huge}}
\def\eH{\end{Huge}}
\def\bL{\begin{Large}}
\def\eL{\end{Large}}
\def\bl{\begin{large}}
\def\el{\end{large}}
\def\beq{\begin{eqnarray}}
\def\eeq{\end{eqnarray}}
\def\beqnn{\begin{eqnarray*}}
\def\eeqnn{\end{eqnarray*}}

\def\bit{\begin{itemize}}
\def\eit{\end{itemize}}
\def\bsc{\begin{screen}}
\def\esc{\end{screen}}

\def\eps{\epsilon}
\def\th{\theta}
\def\del{\delta}
\def\omg{\omega}

\def\e{{\rm e}}
\def\exp{{\rm exp}}
\def\arg{{\rm arg}}
\def\Im{{\rm Im}}
\def\Re{{\rm Re}}

\def\sup{\supset}
\def\sub{\subset}
\def\a{\cap}
\def\u{\cup}
\def\bks{\backslash}

\def\ovl{\overline}
\def\unl{\underline}

\def\rar{\rightarrow}
\def\Rar{\Rightarrow}
\def\lar{\leftarrow}
\def\Lar{\Leftarrow}
\def\bar{\leftrightarrow}
\def\Bar{\Leftrightarrow}

\def\pr{\partial}

\def\>{\rangle} 
\def\<{\langle} 
\def\RR {\rangle\!\rangle} 
\def\LL {\langle\!\langle} 
\def\const{{\rm const.}}

\def\e{{\rm e}}

\def\Bstar{\bL $\star$ \eL}

\def\etath{\eta_{th}}
\def\irrev{{\mathcal R}}
\def\e{{\rm e}}
\def\noise{n}
\def\hatp{\hat{p}}
\def\hatq{\hat{q}}
\def\hatU{\hat{U}}

\def\hatA{\hat{A}}
\def\hatB{\hat{B}}
\def\hatC{\hat{C}}
\def\hatJ{\hat{J}}
\def\hatI{\hat{I}}
\def\hatP{\hat{P}}
\def\hatQ{\hat{Q}}
\def\hatU{\hat{U}}
\def\hatW{\hat{W}}
\def\hatX{\hat{X}}
\def\hatY{\hat{Y}}
\def\hatV{\hat{V}}
\def\hatt{\hat{t}}
\def\hatw{\hat{w}}

\def\hatp{\hat{p}}
\def\hatq{\hat{q}}
\def\hatU{\hat{U}}
\def\hatn{\hat{n}}

\def\hatphi{\hat{\phi}}
\def\hattheta{\hat{\theta}}

\def\iset{\mathcal{I}}
\def\fset{\mathcal{F}}
\def\pr{\partial}
\def\traj{\ell}
\def\eps{\epsilon}
\def\U{\hat{U}}

\def\U{U_{\rm cls}}
\def\P{P_{{\rm cls},\eta}}
\def\traj{\ell}
\def\cc{\cdot}

\def\DZ{D^{(0)}}
\def\Dcls{D_{\rm cls}}

\newcommand{\relmiddle}[1]{\mathrel{}\middle#1\mathrel{}}

\title{Localization and delocalization properties in quasi-periodically 
perturbed Kicked Harper and Harper models
}
\author{Hiroaki S. Yamada}
\affiliation{Yamada Physics Research Laboratory,
Aoyama 5-7-14-205, Niigata 950-2002, Japan}
\author{Kensuke S. Ikeda}
\affiliation{College of Science and Engineering, Ritsumeikan University, 
Noji-higashi 1-1-1, Kusatsu 525-8577, Japan}

\date{\today}
\begin{abstract}
We numerically study the single particle localization and delocalization phenomena of an 
initially localized wave packet in the 
kicked Harper model  (KHM) and Harper model subjected to quasi-periodic perturbation
composed of $M-$modes.
Both models are localized in the monochromatically perturbed case $M=1$. 
KHM shows localization-delocalization transition (LDT) above $M\geq2$ as increase of
 the perturbation strength $\eps$.
In contrast, in a time-continuous Harper model with the perturbation, 
it is confirmed that the localization persists for $M=2$
 and the LDT occurs for $M\geq 3$. 
Furthermore, we investigate the diffusive property of the delocalized wave packet 
 in the KHM and Harper model
 for $\eps$ above the critical strength $\eps_c$ ($\eps>\eps_c$) 
 comparing with other type systems without localization, 
which takes place a ballistic to diffusive transition in the wave packet dynamics 
as the increase of $\eps$.  
\end{abstract}



\maketitle


\section{Introduction}
\label{sec:intro}
The quantum kicked rotor  (KR), representing the dynamics of a periodically
kicked pendulum, is a well studied as example of a classically 
nonintegrable system \cite{casati79,fishman82,reichl04}.
The dynamical localization of the quantum KR has been interpreted through 
a mapping to the Anderson tight-binding model for a particle in a disordered lattice 
\cite{fishman82}.
Furthermore, in the KR systems, 
 a transition from the dynamical localization to delocalization can be 
 caused by second periodic series of kicks, and periodic modulation of the kick amplitude, and so on
\cite{abal02,wang08,schomerus08}.

Similar to the KR systems, the existence of localization-delocalization transition (LDT)
 in the dynamics has been investigated in the kicked Harper model (KHM), 
 whose classical counterpart is a non-integrable system 
\cite{biddle09,molina14,rayanov15,ganeshan15,major18,castro19,alexAn21}.
 There are three types of main dynamical states of the quantum wave packet, localized, normal diffusion, 
 and ballistic spread, corresponding to the change of the potential strength in the KHM. 
 In that respect, it is the same as the Harper model without the kicks.
  In the case of the Harper model, at one point of the potential strength, 
 the  LDT takes place due to the duality of the system
 \cite{harper55,hofstadter76,aubry80}.
  On the other hand, the phase diagram of the localized/delocalized state in KHM 
 has a nested structure and is quite complicated
 \cite{artuso94,prosen01,kolovsky03,levi04}.
In the case of KHM, there is also studies on the LDT caused by 
 the second kick series or kick intensity modulation
 \cite{kolovsky12,wang13,qin14,cadez17,ravindranath21,lakshminarayan03,mishra16,ray18}. 
 
In general, for the periodically
driven Hamiltonian systems, Floquet states represent the
natural generalization of the stationary eigenstates for time-independent
systems.  
 Floquet engineering in the periodically kicked systems is interesting 
because they can exhibit more complex dynamics and
the quantum state is more controllable through external driving in comparison to
their static systems
\cite{kolovsky12,wang13,qin14,cadez17,ravindranath21,lakshminarayan03,mishra16,ray18}.

Recently, the LDT by using the KR has been
experimentally explored using cold atoms in optical lattices
\cite{sadgrove07,dana08,chabe08,lemarie10}. 
It is also feasible to experimentally realize  the dynamical LDT
 through the diffusion of wave packets
in the pulsed 1D incommensurate optical lattice such as the KHM
\cite{sarkar17}.

On the other hand, we have also investigated the characteristics of the 
LDT in the polychromatically perturbed kicked Anderson model  (KAM), 
 and the results that correspond well with those in the KR
 have been obtained \cite{yamada04,yamada10,yamada15,yamada18,yamada20,yamada20a}.
[ Note that in our previous paper, we used the expression Anderson map (AM) 
 for the KAM. ]
The dynamics is not solely determined by the
strength $\eps$ of the perturbation, but also the number of color $M$ of the coherent perturbation.
It was shown that for the time-continuous 1D Anderson model modulated 
by a quasi-periodic time-perturbation of $M$ colors, 
if there are three or more color perturbations ($M\geq 3$), 
the Anderson localized states without the perturbation can be delocalized, and 
the LDT occurs as the increase of $\eps$ \cite{yamada21,yamada22}.

In this work, we study the dynamical LDT
in the KHM and Harper models which are perturbed 
by the quasi-periodic oscillations.
Specifically, in the KHM of $2 \leq M < \infty$, there are LDTs and the  characteristics 
are similar to those of the KAM. 
However, in the Harper model, the case of $M=2$ is completely localized
 on the small side of $\eps$, and tends to persist the localization 
 and the LDT has not been observed even on the large side of  $\eps$. 
In the time-continuous  Harper model with the polychromatic perturbation, 
the LDT does not occur in the case of $M=2$, 
and the LDT occurs only in a case of $M\geq 3$, as seen in the case 
of the 1D Anderson model  with the polychromatic perturbation \cite{yamada21,yamada22}.

It should be noticed that the persistence of 
the localization for $1 \leq M < \infty$ 
is mathematically claimed in the regime of weak 
enough dynamical perturbations and strong disorder potential 
\cite{Soffer03,Bourgain04,hatami16}.
In this regard,  the present paper reports the details of novel results on dynamical delocalization
in the perturbed Harper model. 
Of particular interest to us is dependence of the LDT on the  
 two parameters of coherent polychromatic perturbation, i.e., 
 number of the colors $M$ and perturbation strength $\eps$.

Furthermore, by using the other type system without localization,
which is related to  both the KHM and Harper models, 
we report the change in the wave packet dynamics
 from the ballistic spreading in the unperturbed Bloch state
 to the normal diffusion when the perturbation strength is increased. 
Hereinafter, this transition is referred to as a ballistic-diffusive transition (BDT) 
of wave packet spreading. 
The timescale for the BDT 
depends on the strength $\eps$ as well as $M$.
 In the next section, those indicating BDT are introduced as B-type system. 
 In contrast, those indicating LDT are used as A-type system.
 
 In Sec.\ref{sec:model} we introduce the model and the basic property. 
Section \ref{sec:LDT} explores the dynamical LDT 
when  perturbed by quasi-periodic oscillation with the component $M$ 
and the strength $\eps$.
 In contrast, in Sect.\ref{sec:ballistic}, we focus on the diffusive property of the delocalized states,
and show the relation to BDT by  the coherent perturbation in the B-type system. 
In Sect.\ref{sect:typeAB}, 
the relationship of the diffusive states in the A-type and B-type systems is shown.
In the last section, we 
summarize our findings while comparing them with other results.

\section{Models and some preliminaries}
\label{sec:model}
We deal with  the dynamically perturbed kicked Harper model (KHM),
\beq
H_{kick}(t)&=&\sum\limits_{n=1}^N {|n \rangle V(n)[L+f_\eps(t)] \langle n|} \delta_1(t) \nn \\
 &+& T\sum\limits_{n}^N ({|n \rangle \langle n+1|+
|n+1 \rangle \langle n|}), 
\label{eq:Hamiltonian}
\eeq
where $\delta_1(t)=\sum_{m\in {\Bbb Z}}\delta(t-m)$.
The on-site energy sequence is 
\beq
  V(n)=2V \cos(2\pi Q n), 
\eeq
where $\{|n \rangle \}$ is an orthonormalized basis set and the $Q$ is an irrational number. 
 $V$ is potential strength, and $T$ denotes the hopping energy 
 between adjacent sites, respectively.
We tale $Q=\frac{\sqrt{5}-1}{2}$ and $T=-1$ throughout the present paper.

The coherently time-dependent part $f_\eps(t)$ is given as,
 \beq
  f_\eps(t)=\eps f(t)=\frac{\eps}{\sqrt{M}} \sum_i^M\cos(\omega_i t+\theta_i), 
\eeq
where $M$ and $\eps$ are the number of frequency components and 
the strength of the perturbation, respectively.
Note that the long-time average of the total power of the perturbation is normalized to 
$\overline{f_\eps(t)^2}=\eps^2/2$ and $\{\theta_i \}$ are the initial phases.
Since we see the long-time behavior regardless of how the initial phases are taken, 
we basically set it as $\{\theta_i=0 \}$, but we also  partially  deal with the case of 
 random phases. 
The frequencies $\{ \omega_i\}(i=1,...,M)$ are taken as mutually incommensurate 
numbers of order $O(1)$.
Following two types of cases: 
\beq
\begin{cases}
L=1 & ({\rm A-type~system})\\
L=0 & ({\rm B-type~system}).
\end{cases}
\eeq
behave completely differently to the perturbation.
In the  A-type system the LDT occurs by increasing of 
the perturbation strength $\eps$.
Even in the  B-type system the normal diffusion occurs by increasing of  $\eps$
through the BDT.

For the initial wave packet $<n|\Psi(t=0)>=\delta_{n,n_0}$ localized at the site $n_0$, 
we calculate the time evolution of the wavefunction $|\Psi(t)>$ 
using  Schrodinger equation: 
\beq
i\hbar\frac{\pr |\Psi(t)>}{\pr t}=H(t)|\Psi(t)>.
\eeq
We monitor the spread of the wave function in the site space by the 
mean square displacement (MSD),
\beq
m_2(t) = \sum_{n}(n-n_0)^2 \left< |\phi(n,t)|^2 \right>, 
\eeq
where $\phi(n,t)=<n|\Psi(t)>$ is the site representation of the wave function.


In the numerical simulation for the time-continuous system, 
we used  2nd order symplectic integrator with time step $\Delta t=0.02 \sim 0.05$.
The number of steps is $10^5\sim10^7$.
We mainly use the system size $N=2^{14}-2^{16}$, 
and $\hbar=1/8$ for KHM,  and $\hbar=1$ and  $\hbar=1/8$ for Harper model.

\begin{figure}[htbp]
\begin{center}
\includegraphics[width=6.2cm]{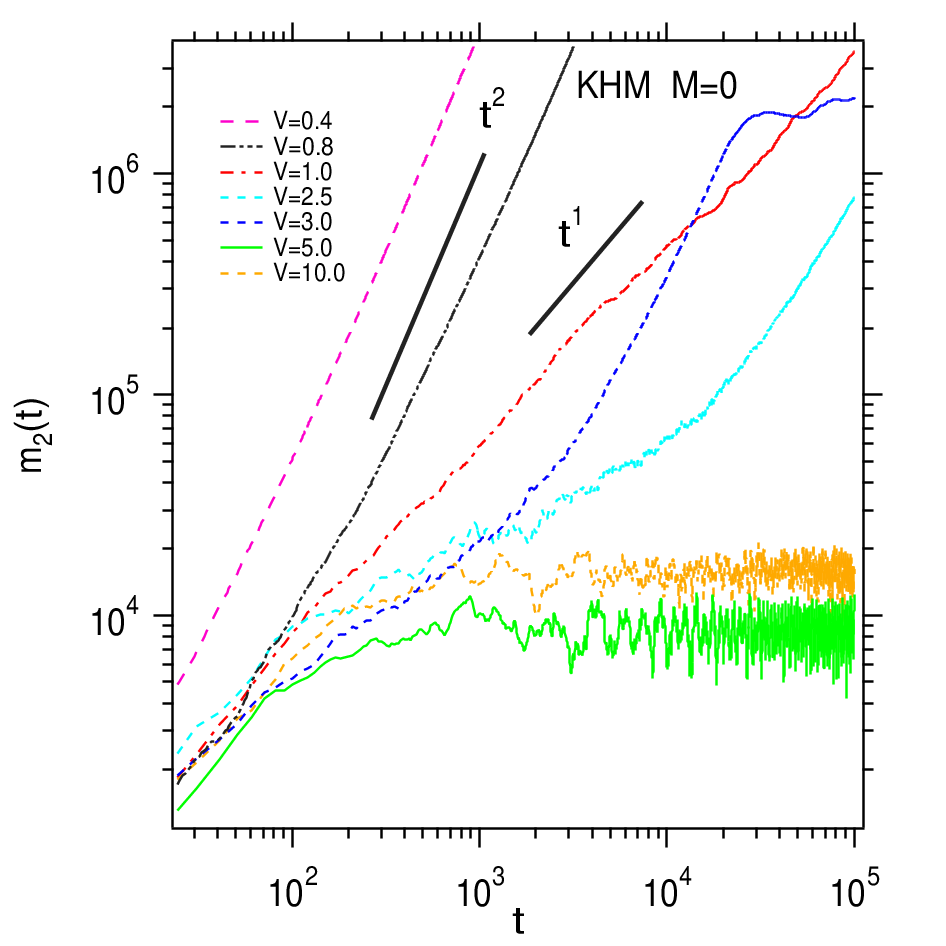}
\caption{(Color online)
The double logarithmic plots of $m_2$ as a function of $t$ 
for some values of the potential strength $V=0.4,0.8,1.0,2.5,3.0,5.0,10.0$
in the unperturbed  A-type KHM ($\eps=0$). $\hbar=1/8$.
The solid lines have slope 1 and 2.
}
\label{fig:KHM0}
\end{center}
\end{figure}

The unperturbed KHM ($\eps=0$) is known to 
take a localized, critical, and extended state with a change of $V$. 
Figure \ref{fig:KHM0} shows the time-deprendence of the MSD $m_2(t)$
when the potential strength $V$ is changed from the localized side 
to the delocalized one. 
In all cases, the  $m_2(t)$ grows $m_2 \sim t^2$ for small $t$, 
but the behavior for $t>>1$ is quite different with the parameter $V$.
If $V$ is small, we can see ballistic spreading $m_2 \sim t^2$. 
When $V=1$, normal diffusive behavior close to $m_2 \sim t^1$ is observed, 
and when $V$ is large, it tends to be localized. 
However, the $V-$dependence in the KHM is considerably more complicated 
than the unperturbed Harper model introduced below as a time-continuous system.
Indeed, even if it is localized, the $V-$dependence of the localization length 
cannot be simply given diffrent from the case of Harper model. 
In the corresponding classical dynamics, the unperturbed KHM shows chaos, 
but the unperturbed Harper model is an integrable system in a classical limit. 
The eigenvalue problem of the periodically kicked system can be replaced 
with the eigenvalue problem of the tight-binding system by the Maryland transform.
 As given in  Appendix \ref{app:maryland}, one can easily check that
the eigenvalue problem of the quantum map system interacting with 
$M$-color modes can be transformed into $d(=M+1)$-dimensional
lattice problem with disorder by Maryland transform. 
Without loss of generality, mainly for a fixed value $V=5$ to be localized 
we investigate the LDT in the A-type systems with increasing of  $M$ and $\eps$ 
of the quasi-periodic perturbation  in the next section. 


Furthermore, we examine the LDT for 
 a time-continuous system in which $\delta_1(t)$ is replaced by 1 in the Hamiltonian $H_{kick}(t)$. 
 This model was introduced as an model for electron 
 in a two-dimensional crystal in an external  magnetic field, and we call it 
 Harper model in this paper.
[There are also references that describes this model as Aubry-Andre model 
or Aubry-Andre-Harper model. 
]
It can be said that the KHM system is a map version of the Harper model.

The nature of the unperturbed Harper model ($\eps=0$) has long been well studied 
physically and mathematically. 
It has been also shown that the LDT exists in the unpertubed Harper model  due to  the duality 
\cite{hiramoto88,geisel91,wilkinson94}.
For $V >1$ all eigenstates are localized and the spectrum is pure point. 
For $V >1$ the localization length $\xi$ is given $\xi =\frac{1}{\log V}$ 
independent on the energy. 
In this case, the increase in $V$ causes a monotonous decrease in $\xi$. 
For $V <1$ the states are extended and the spectrum is absolutely continuous.
For the critical value  $V=V_c=1$  eigenstates are critical and 
the spectrum is singular continuous. 
It is also numerically suggested that spread of an initially localized wavepacket 
is localized for $V >1$, and is ballistic for $V <1$
and is diffusive for $V=1$, respectively
\cite{math-comment}.

\begin{figure}[htbp]
\begin{center}
\includegraphics[width=6.0cm]{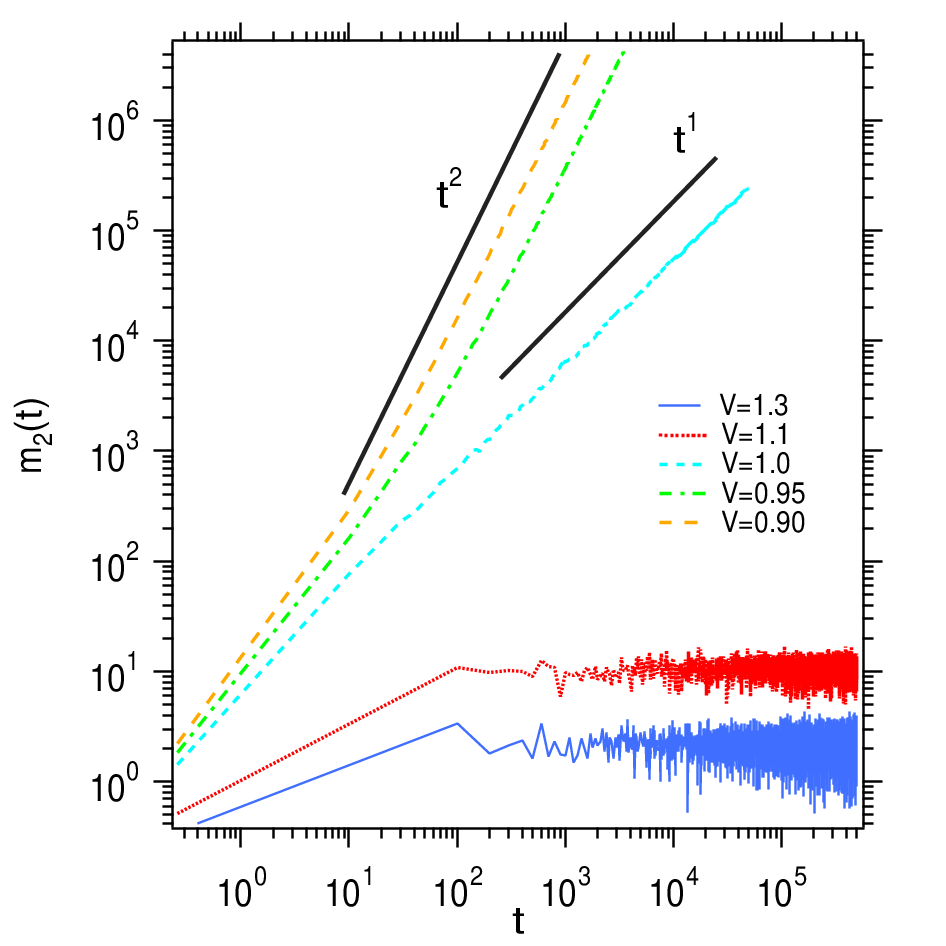}
\caption{(Color online)
The double logarithmic plots of $m_2$ as a function of $t$ 
for some values of the potential strength $V=1.3, 1.05, 1.0, 0.95, 0.9$
in the unperturbed  A-type Harper model ($\eps=0$). $\hbar=1$.
The solid lines have slope 1 and 2.
}
\label{fig:harper0}
\end{center}
\end{figure}

Figure \ref{fig:harper0}  shows the MSD in the unperturbed Harper model 
when the potential strength $V$ is changed 
from the localized side ($V>1$) to the delocalized side ($V <1$).
Although there are fluctuations, it changes from $m_2 \sim t^0$ in the case of 
localization to a ballistic spread of $m_2 \sim t^2$ via normal diffusive behavior $m_2 \sim t^1$ 
in the case of the critical state ($V=1$). 
When $ V<1$, it is ballistic regardless of the value of $V$. 
That is, 
\beq
m_2(t) \sim 
\begin{cases}
t^0  {\rm (localization)} & V > 1  \\
t^{1}  {\rm (normal~diffusion)} & V=V_c=1  \\  
t^{2}  {\rm (ballistic~spreading)} &V<1
\end{cases}
\eeq
Using the  A-type Harper model, we fixed at $V=1.3$ as an unperturbed localized side, and
investigate  the LDT  by imposing $f_\eps(t)$,  with 
 comparing with results in the already reported 1D kicked Anderson  
 and the Anderson models \cite{yamada20a,yamada21,yamada22}. 

In general, anomalous diffusion $m_2(t) \sim t^\alpha$
 characterized by diffusion index $\alpha$ is expected 
 for the LDT  ($0<\alpha<1$)  and BDT ($1<\alpha<2$).
The instantaneous diffusion index $\alpha_{ins}(t)$ is also used 
to directly investigate the existence of the LDT and the BDT:
\beq
 \alpha_{ins}(t)=\frac{d\log m_2(t)}{d\log t}.
\eeq
In the A-type system, 
above the critical point $\eps>\eps_c$ the index becomes unity 
($\alpha=1$) indicating the normal diffusion,
and for $\eps<\eps_c$ it decreases to zero ($\alpha=0$)  indicating localization.
Around the critical point $\eps \simeq \eps_c$  for LDT
 $\alpha_{ins}(t) \simeq \alpha_c (<1)$ for $t>>1$.
 In the B-type system, 
above $\eps>\eps_b$ the $\alpha(t)$ becomes unity, 
and for $\eps<\eps_b$ it increases to 2 indicating the ballistic spreading.

\section{Dynamical Localization-Delocalization Transition}
\label{sec:LDT}
In this section, we  examine the dynamical property by changing the 
parameters $M$ and $\eps$ of the perturbation $f_\eps(t)$ in the  A-type system
of the KHM with $V=5$ and Harper model with $V=1.3$, respectively.
 
\subsection{Kicked Harper model: A-type case}

\begin{figure}[htbp]
\begin{center}
\includegraphics[width=4.4cm]{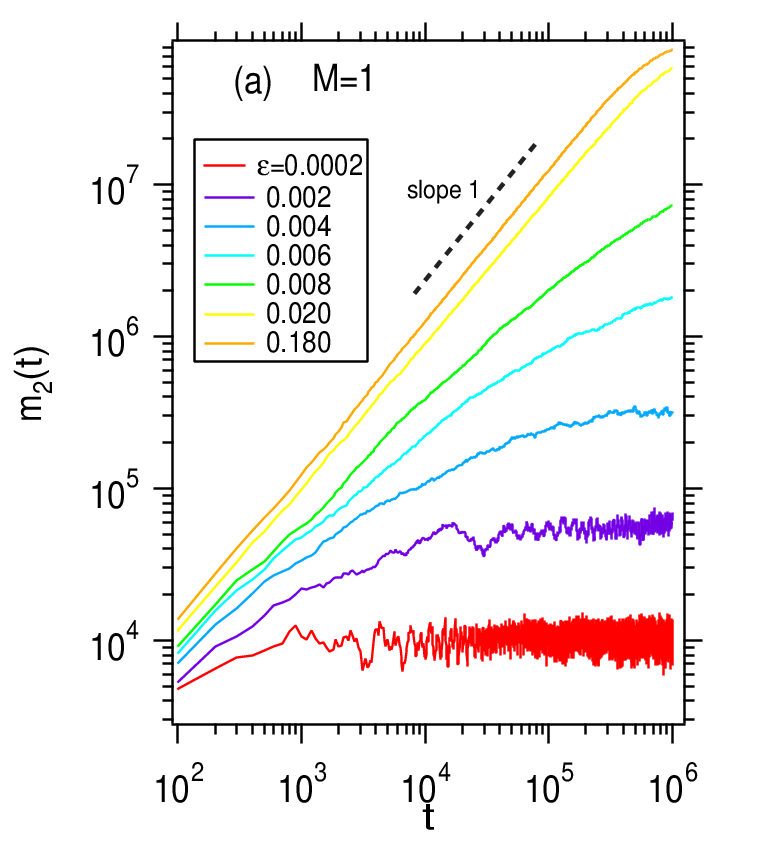}
\hspace{-5mm}
\includegraphics[width=4.4cm]{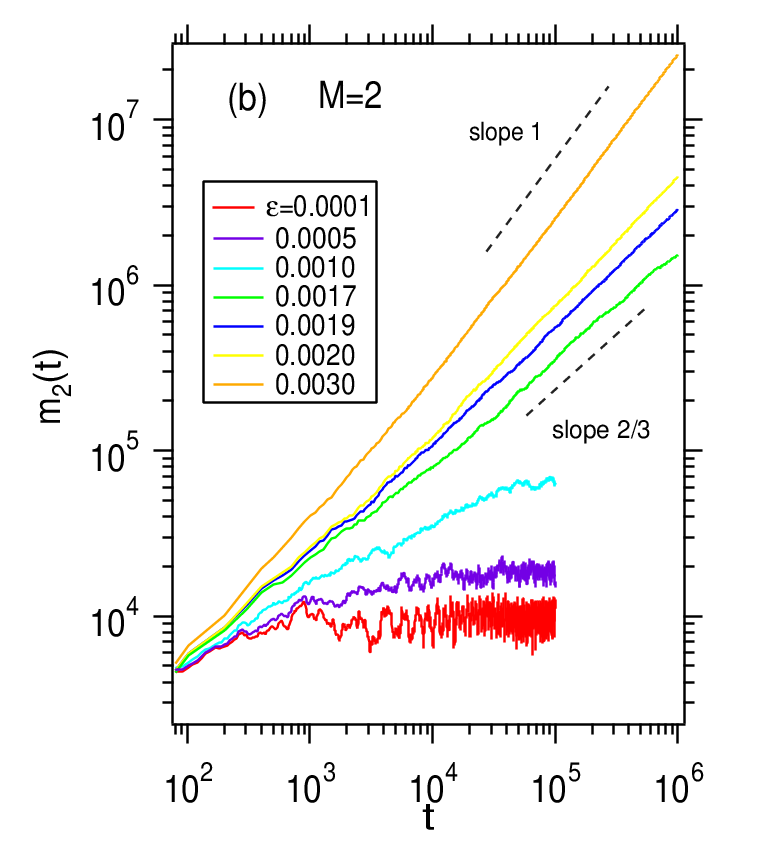}
\hspace{5mm}
\includegraphics[width=4.4cm]{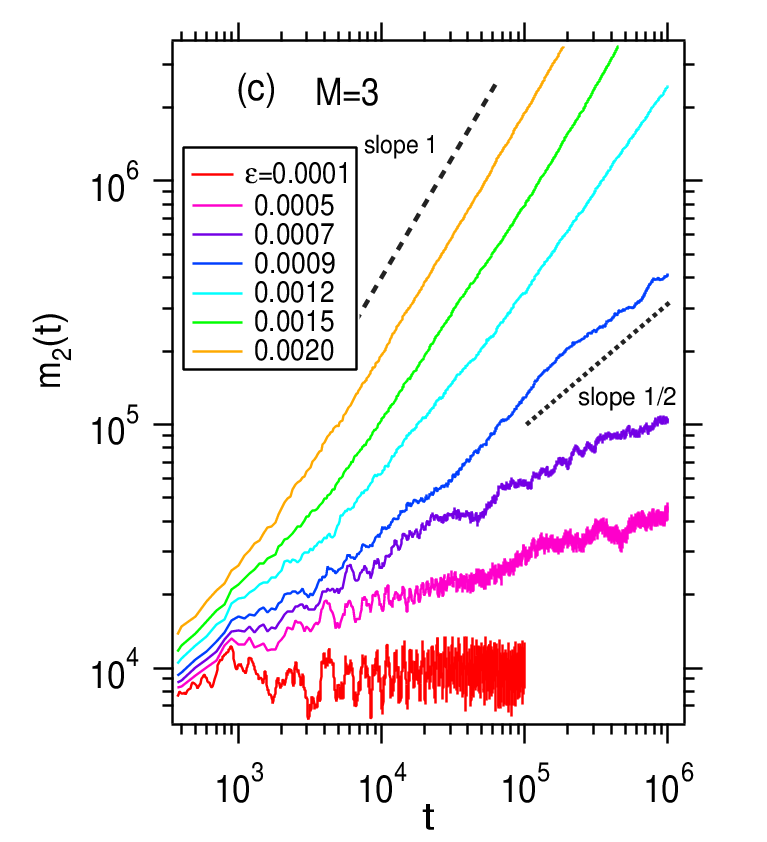}
\hspace{-5mm}
\includegraphics[width=4.4cm]{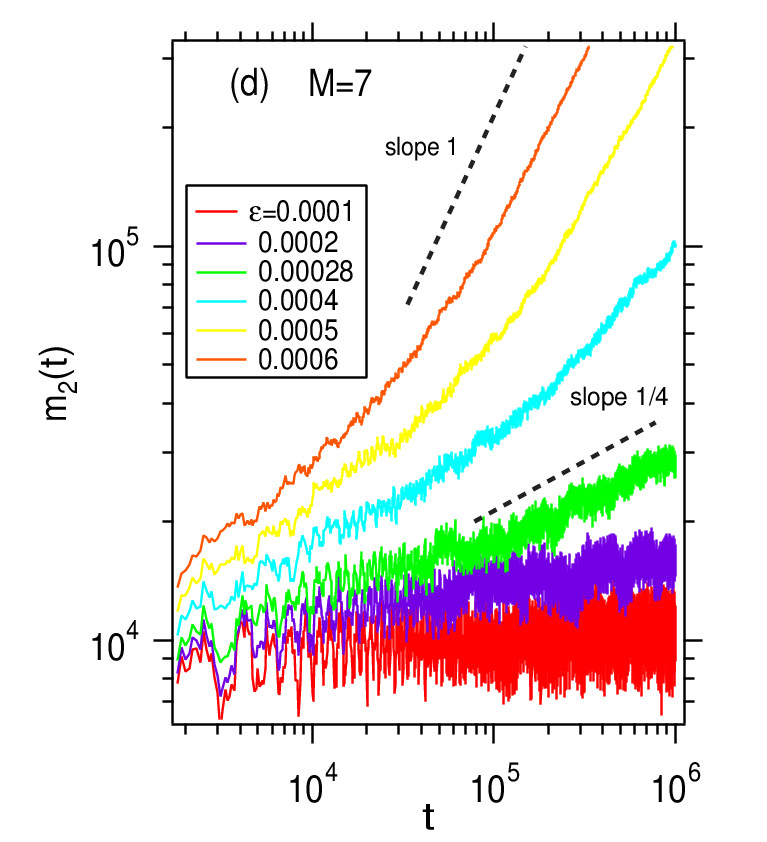}
\caption{(Color online) \label{fig:t-map-c1c3-msd}
The double-logarithmic plots of $m_2(t)$ as a function of time
for different values of the perturbation strength $\eps$
in  the polychromatically perturbed  A-type 
KHM of $V=5$ with (a)$M=1$, (b)$M=2$, (c)$M=3$, (d)$M=7$. $\hbar=1/8$.
The values of $\eps$ used are $\eps=0.0002,0.0020,0.0040,0.0080,0.0200,0.1800$ 
from bottom to top in the panel (a),
$\eps=0.0001,0.0005,0.0010,0.0017,0.0019,0.0020,0.0030$ from bottom to top in the panel (b), 
$\eps=0.0001,0.0005,0.0007,0.0009,0.0012,0.0015,0.0020$ from bottom to top in the panel (c),
and $\eps=0.0001,0.0002,0.00028,0.0004,0.0005,0.0006$ from bottom to top in the panel (b), respectively.
The dashed lines have slope 1  in the panel (a), slope 1 and 2/3 in the panel (b), 
 slope 1 and 1/2 in the panel (c), and slope 1 and 1/4 in the panel (d), respectively. 
}
\end{center}
\end{figure}

Figure \ref{fig:t-map-c1c3-msd}(a) shows the time-dependence of MSD 
 in the log-log plots when $M=1$.
When $\eps$ is small, it is clearly completely localized.
Even when $\eps$ increases, 
it spreads diffusively within the initial time,
 but as time elapses, the $t-$linear growth begins to wither and the wave packet tends to 
 be localized. As a result, no transition to the delocalized state is seen
 in the monochromatically perturbed case. 

In Fig.\ref{fig:t-map-c1c3-msd}(b)(c)(d), the MSD are shown for $M\geq2$.
They are localized when $\eps$ is small, but the LDT occurs with 
 a certain critical value $\eps_c$, and $\eps$ becomes larger than $\eps_c$,  
 warps upward (upward deviation) can be seen  in the double-logarithmic plots.
For $\eps>\eps_c$ the normal diffusive behavior $m_2 \sim t^1$ appears as $t \to \infty$.

As shown in Fig.\ref{fig:t-map-c1c3-msd}(b)(c)(d), around $\eps=\eps_c$ 
the time-dependence of  MSD can be approximately described 
by the sub-diffusive spreading
\beq
  m_2(t) \sim t^\alpha,  \alpha \simeq \frac{2}{M+1}, 
\eeq
depending on $M$.

\begin{figure}[htbp]
\begin{center}
\includegraphics[width=6.5cm]{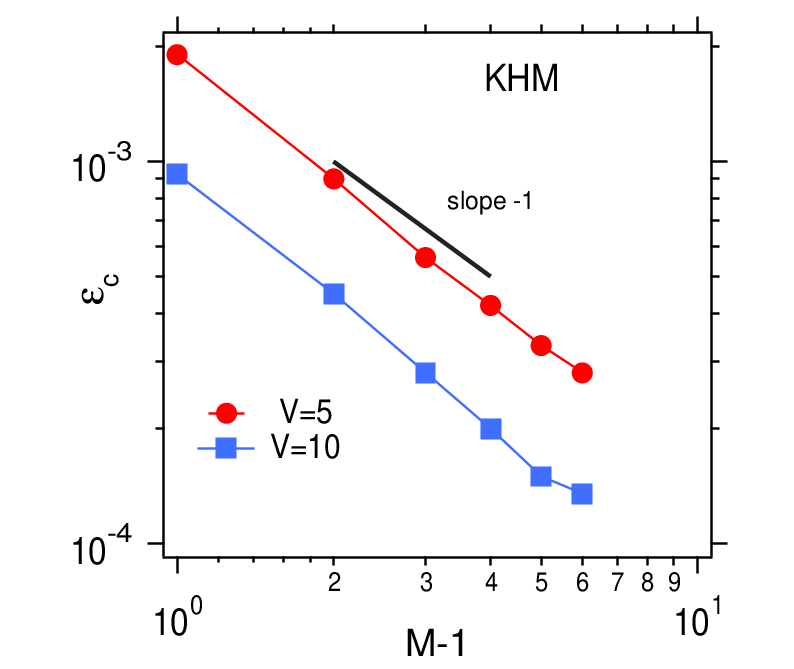}
\caption{\label{fig:V5epsc-M}(Color online)
The critical perturbation strength $\eps_c$ as a function of  $(M-1)$ 
for  A-type KHM with $V=5$ and $V=10$. 
The black solid line shows $\eps_c \propto 1/(M-1)$.
}
\end{center}
\end{figure}

As shown in Fig.\ref{fig:V5epsc-M}, for $M\geq 2$ the $M-$dependence of the 
critical strength $\eps_c$ indicates the inverse power-law 
\beq
 \eps_c \propto \frac{1}{V}\frac{1}{(M-1)}.
\eeq
The same $M-$dependence even for  $V=10$ can be obtained,
as seen in Fig.\ref{fig:V5epsc-M}.
This difference in $\eps_c$ due to $V$ can be interpreted by the Maryland transform
 in Appendix \ref{app:maryland}. 
According to the Maryland transform, the effect of $V$ in the diagonal term is saturated 
in the region $V>V^*(\equiv 0.38)$, and 
 the nature of the off-diagonal term depends on $\eps V$, so the critical value $\eps_c$ 
  for $V=10$ can be interpreted as a half of the critical value $\eps_c$
  of the transition point for $V=5$. 


It can also be seen that for $\eps>\eps_c$, $m_2(t)$ asymptotically becomes 
to normal diffusion as $t \to \infty$.
Therefore, we can say that the quasi-periodically driven system also may delocalize 
even in the absence of coupling with its environment.

\begin{figure}[htbp]
\begin{center}
\includegraphics[width=4.4cm]{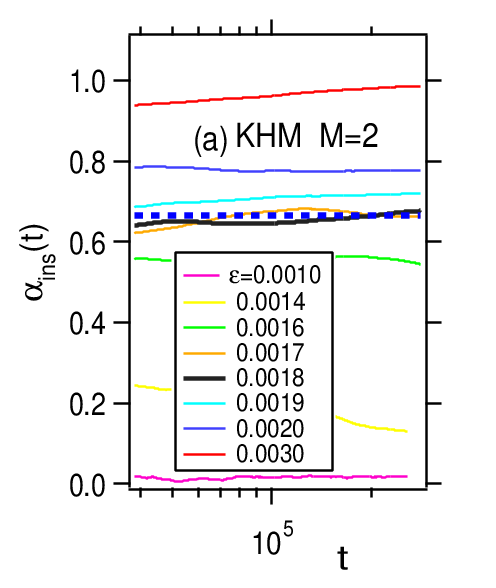}
\hspace{-5mm}
\includegraphics[width=4.4cm]{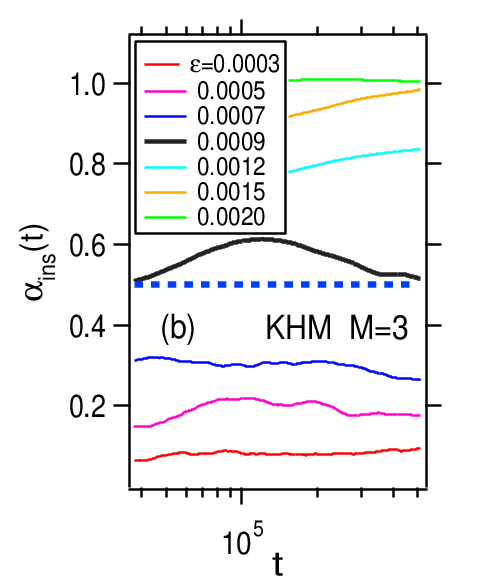}
\caption{(Color online) \label{fig:A-KHM-c2c3}
The time-dependence of $\alpha_{ins}(t)$ 
for various strength $\eps$ in the  perturbed A-type KHM of $V=5$ 
with (a)$M=2$ and  (b)$M=3$. $\hbar=1/8$.
The values of $\eps$ used are 
$\eps=0.001, 0.0016, 0.0017, 0.0018, 0.0019, 0.002, 0.003, 0.004$ 
from top to bottom in the panle (a),
and  $\eps=0.0003, 0.0005, 0.0007, 0.0009, 0.0012, 0.0015, 0.0020$ 
from top to bottom in the panle (b), respectively.
The dotted lines indicate $\alpha=2/(M+1)$.
}
\end{center}
\end{figure}

Using $\alpha_{ins}(t)$, we confirm the above LDT trend.
As shown in Fig.\ref{fig:A-KHM-c2c3}, 
in the case of KHM, we can see that with an increase of $\eps$
the tendency of $\alpha_{int} \to 0$ changes as  $\alpha_{int} \to 1$ 
through $\alpha_{int}(t)=\alpha_c \simeq 2/(M+1)$ around  $\eps \simeq \eps_c$
for $t>>1$.

\subsection{Harper model: A-type case}

\begin{figure}[htbp]
\begin{center}
\includegraphics[width=4.4cm]{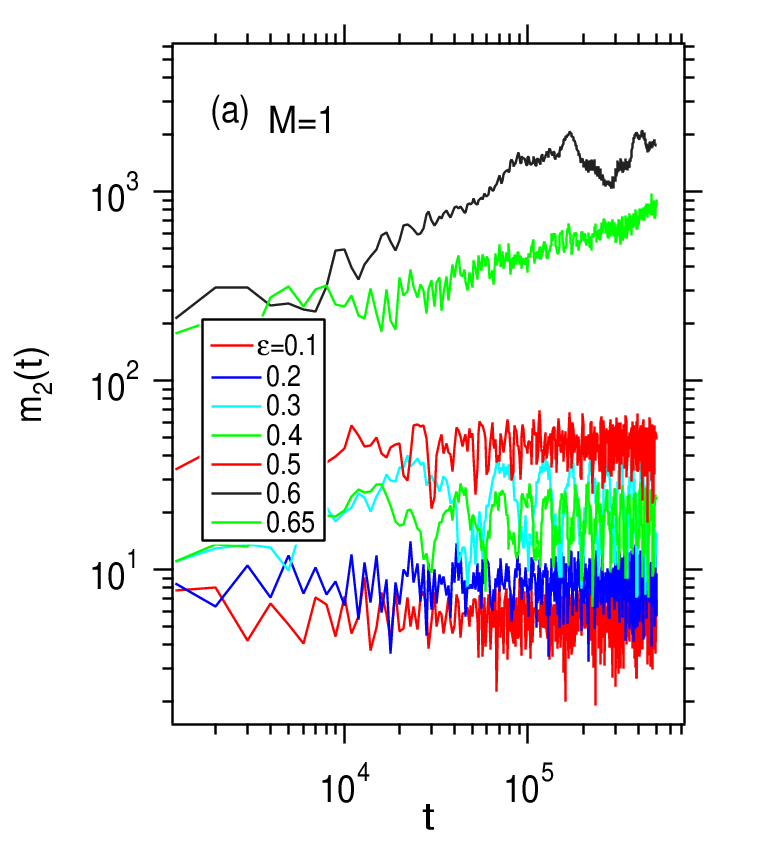}
\hspace{-5mm}
\includegraphics[width=4.4cm]{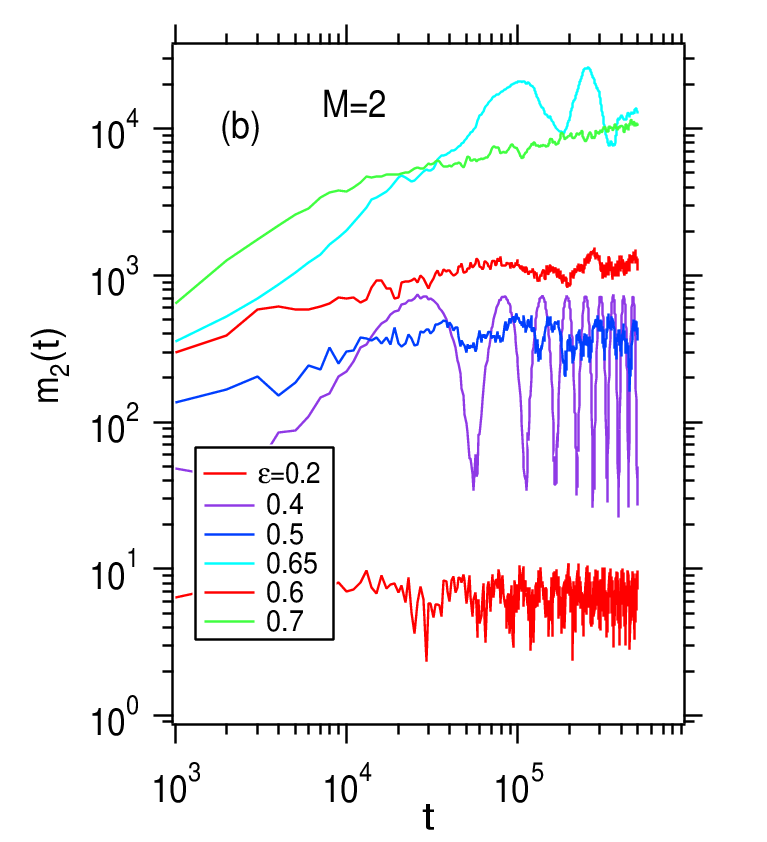}
\caption{(Color online)\label{fig:Hc1-1}
The double logarithmic plots of $m_2$ as a function of $t$ 
for some values of the perturbation strength $\eps$
in the  perturbed A-type Harper model of the potential strength $V=1.3$.
(a)$M=1$ and (b)$M=2$. $\hbar=1$.
The values of $\eps$ used are $\eps=0.1,0.2,0.3,0.4,0.5,0.6,0.65$ from bootom to top in the panel (a) 
and $\eps=0.2,0.4,0.5,0.6,0.65,0.7$ from bootom to top in the panel (b), respectively.
}
\end{center}
\end{figure}

Is the LDT as seen  in the perturbed KHM 
observed when using the Harper model? 
The potential strength is fixed at $V=1.3$, and the time-dependence of the MSD 
for the cases of $M=1$ and $M=2$  is shown in Fig.\ref{fig:Hc1-1}.
Obviously, if $\eps$ is small, it is  localized, and the localization persists 
even for $\eps \sim 0.6$ beyond the region that can be regarded as a perturbation.
It is completely localized in the sense that $D=\frac{m_2}{t} \to 0$ can be guessed.
Also for $M=2$, at least clear subdiffusion does 
not appear even if $\eps$ is increased to $\eps \geq 0.65$. (See Fig.\ref{fig:Hc1-1}(b).)
As a result, it also tends to be localized even for $M=2$, which is also the same as 
the result for the 1D Anderson model with a random sequence as $V(n)$ \cite{yamada22}.

\begin{figure}[htbp]
\begin{center}
\includegraphics[width=4.4cm]{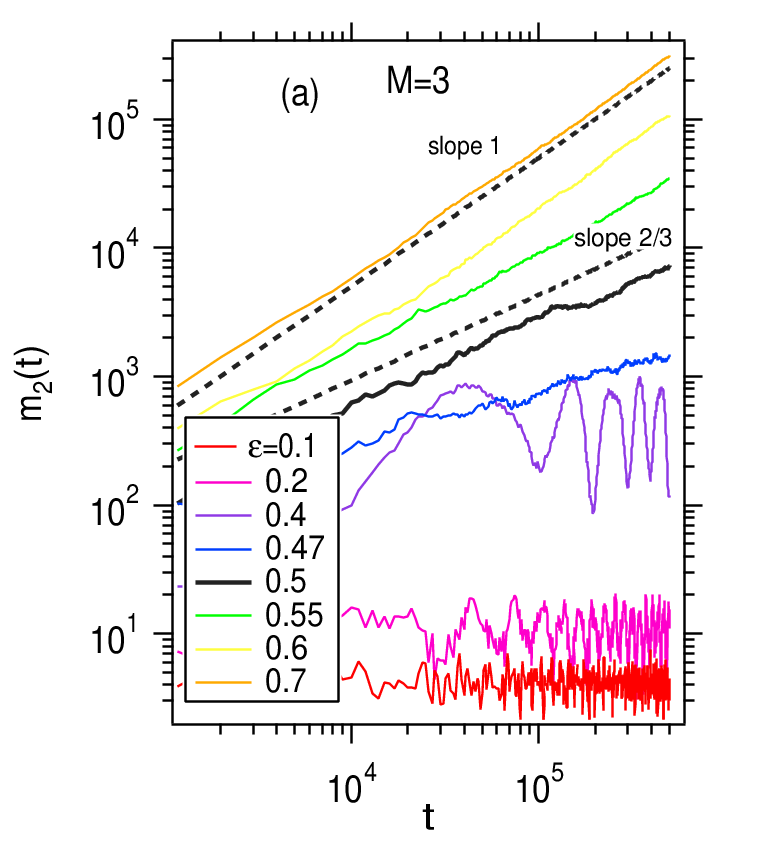}
\hspace{-5mm}
\includegraphics[width=4.4cm]{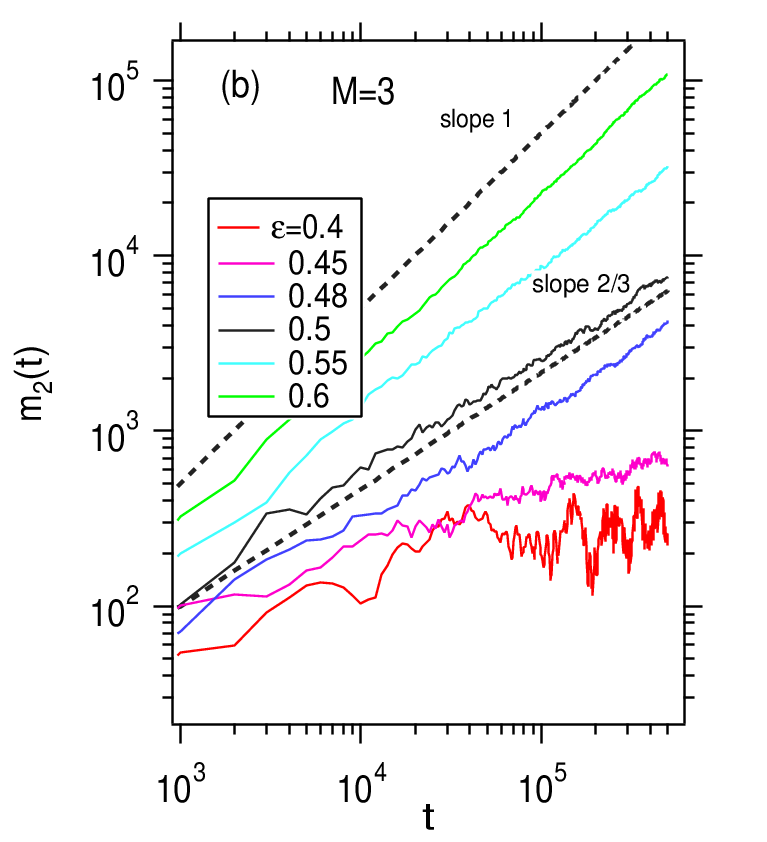}
\hspace{5mm}
\includegraphics[width=4.4cm]{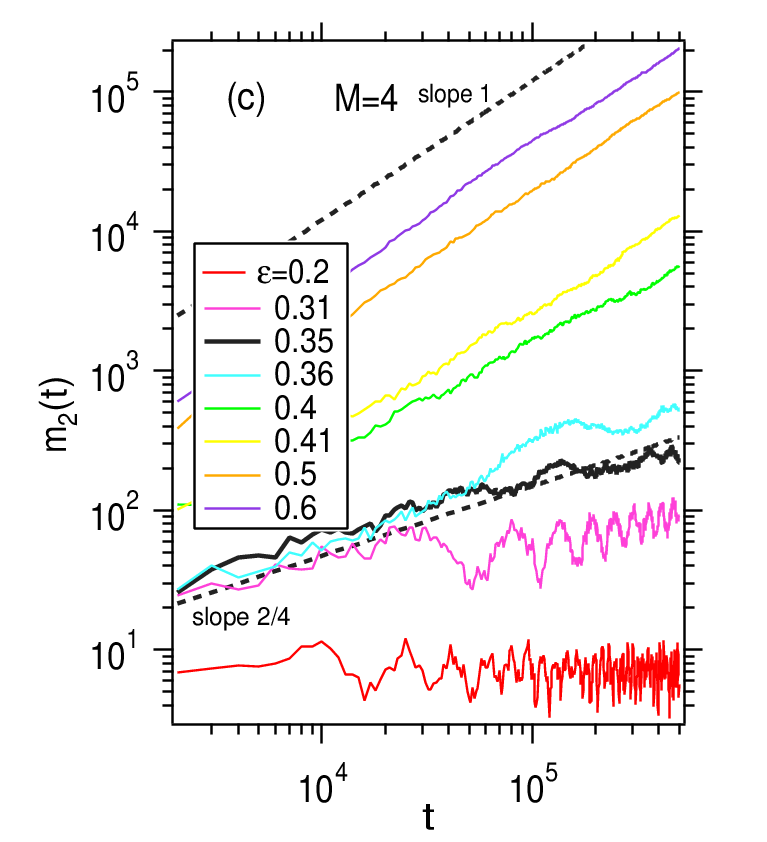}
\hspace{-5mm}
\includegraphics[width=4.4cm]{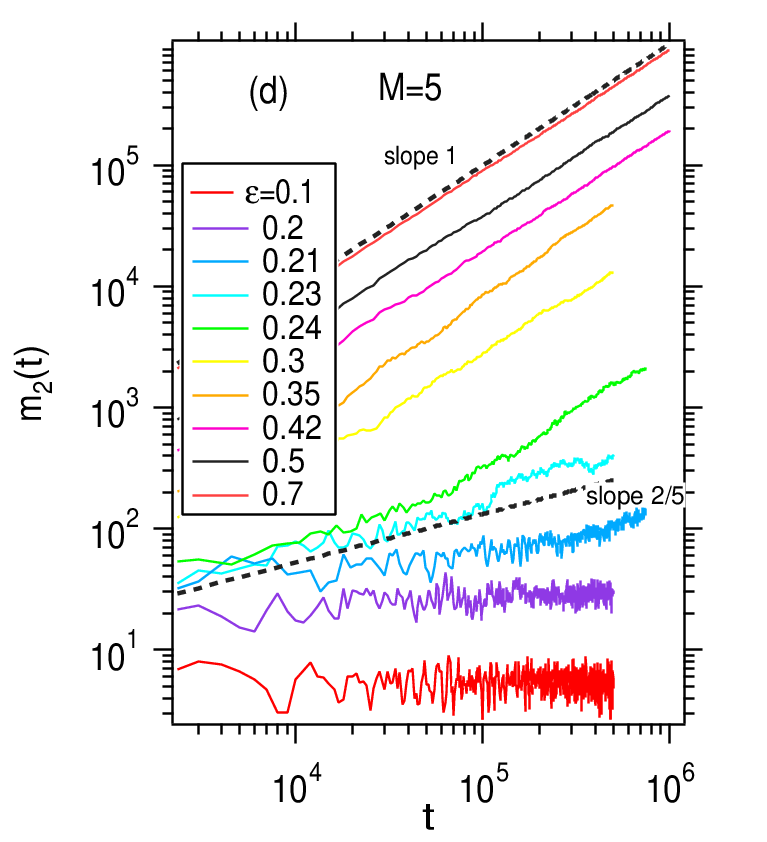}
\caption{(Color online)\label{fig:Hc3-1}
The double logarithmic plots of $m_2$ as a function of $t$ 
for some values of the perturbation strength $\eps$
in the  perturbed A-type Harper model of the potential strength $V=1.3$
with (a)$M=3$,$\{\theta_i\}_0$, (b)$M=3$, $\{\theta_i\}_{random}$, 
(c)$M=4$, $\{\theta_i\}_0$, (d)$M=5$, $\{\theta_i\}_0$. 
$\hbar=1$.
The values of $\eps$ used are $\eps=0.1,0.2,0.4,0.47,0.5,0.55,0.6,0.7$ 
from bottom to top in the panel (a),
$\eps=0.4,0.45,0.48,0.5,0.55,0.6$ from bottom to top in the panel (b), 
$\eps=0.2,0.31,0.35,0.36,0.4,0.41,0.5,0.6$ from bottom to top in the panel (c),
and $\eps=0.1,0.2,0.21,0.23,0.24,0.3,0.35,0.42,0.5,0.7$ from bottom to top in the panel (b), respectively.
$\{\theta_i\}_0$ maens to take all phases to zero, and 
$\{\theta_i\}_{random}$ means that random phases are selected.
The dashed lines have slope 1 and 2/3 in the panel (a) and (b), 
 slope 1 and 2/4 in the panel (c), and slope 1 and 2/5 in the panel (d), respectively. 
}
\end{center}
\end{figure}

Figure \ref{fig:Hc3-1} shows  the result of $m_2(t)$ for $M\geq3$
when $\eps$ is increased. 
First, the results for $M=3$ with different parameters 
are given in Fig.\ref{fig:Hc3-1}(a) and (b).
As seen in Fig.\ref{fig:Hc3-1}(a), 
it is clearly localized in the region $\eps <0.5$, but it changes for $\eps>0.7$, 
and appears to be asymptotic to diffusive one ($m_2 \sim t^1$). 
Figure \ref{fig:Hc3-1}(b) is the result by using the random initial phases $\{ \theta_i\}$ 
of $f_\eps(t)$. 
It is a similar result to the Fig.\ref{fig:Hc3-1}(a). If $\eps$ is small, it is localized, but if $\eps \sim 0.5$, 
the fluctuation becomes large, and if $\eps>0.65$, it becomes diffusive
$m_2 \sim t^1$. 
As a result, in the perturbed A-type Harper model of $M=3$, we can see 
a transition from a localized to a delocalized dynamics 
takes place  through subdiffusion $m_2 \sim t^{2/3}$ 
around the critical value $\eps_c \simeq 0.5$.

Figure \ref{fig:Hc3-1}(c) and (d) show the result of $M=4$ and $M=5$, respectively.
It is also seen that the change from the localized state to 
the normal diffusion as in the case of $M=3$.
In the case of $M=4$, it is clearly localized for $\eps<0.3$, 
and it becomes diffusive at least for $\eps>0.4$. 
In the case of $M=5$, it is clearly localized for $\eps<0.2$, 
and it becomes diffusive at least for $\eps>0.4$. 

From the whole, the transition point $\eps_c$ tends to decrease with $M$. 
However, unlike the case of the KHM, 
a clear subdiffusion ($m_2 \sim t^\alpha$, $0<\alpha<1$)  cannot be detected 
during the transition from the localized ($m_2 \sim t^0$) to delocalized ($m_2 \sim t^1$) 
dynamical property by increasing $\eps$. 
The overall tendency corresponds well to the case of KAM and 
Anderson model where $V(n)$ is random sequence.

\begin{figure}[htbp]
\begin{center}
\includegraphics[width=4.4cm]{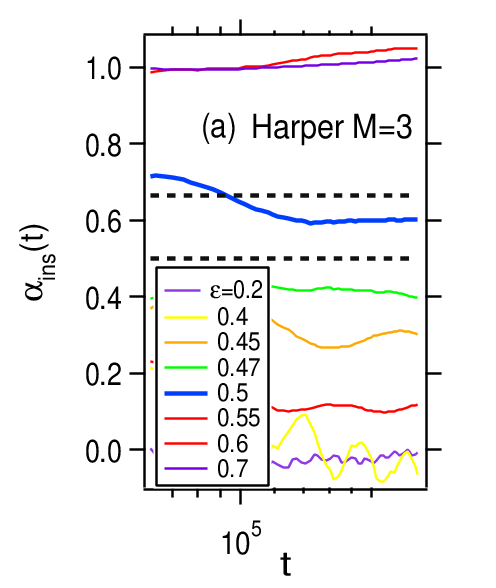}
\hspace{-5mm}
\includegraphics[width=4.4cm]{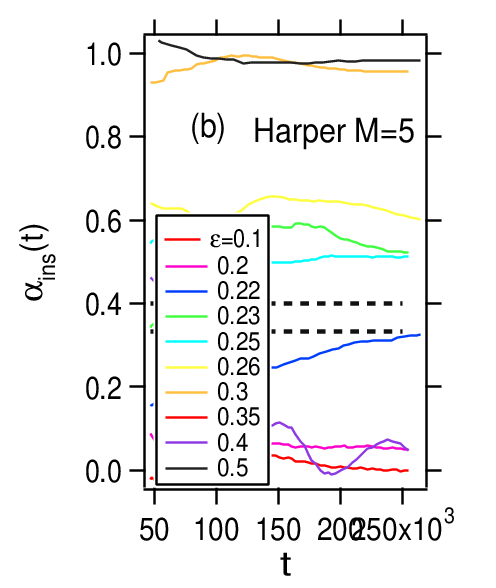}
\caption{(Color online) \label{fig:A-Haper-c3c5}
The time-dependence of $\alpha_{ins}(t)$ 
for various strength $\eps$ in the perturbed  A-type Harper model of $V=1$ 
with (a)$M=3$ and  (b)$M=5$. $\hbar=1$.
The values of $\eps$ used are $\eps=0.20, 0.40, 0.45, 0.47, 0.50, 0.55, 0.60, 0.70$ 
from top to bottom in the panel (a),
and  $\eps=0.1, 0.2, 0.23, 0.25, 0.26, 0.3, 0.35, 0.4, 0.5$ 
from top to bottom in the panel (b), respectively.
The dotted lines indicate $\alpha=2/M$ and $\alpha=2/(M+1)$.
}
\end{center}
\end{figure}

The above trends can be seen on the behavior of  $\alpha_{ins}(t)$.
As shown in Fig.\ref{fig:A-Haper-c3c5}, 
for the Harper model 
the transition from $\alpha \to 0$ to $\alpha \to 1$ exists
with increasing $\eps$, but it is difficult to obtain a stable subdiffusion 
with $\alpha=2/M$ or $\alpha=2/(M+1)$  in the process.

\section{Transition from ballistic spreading to normal diffusion}
\label{sec:ballistic}

In this section, we study the characteristics of the wave packet spreading in 
the  B-type KHM and Harper model ($L=0$) with the coherent oscillation $f_\eps(t)$. 
If $\eps=0$, the potential term disappears, so both KHM and Harper model 
are periodic systems, and the localized wave packet spreads ballistically: 
\beq
m_2(t) \sim t^2.
\eeq
In appendix \ref{app:maryland} 
 for the KHM, we can also see that the diagonal term 
in the Maryland transformed tight-binding model does not depend on the site $n$, 
so it becomes a ballistic motion. 
In the  B-type cases, $V$ and $\eps$ play the same role, so basically fix it as $V=1$ and
we investigate the wave packet dynamics while changing $M$ and $\eps$ of the coherent oscillation part.

\subsection{Kicked Harper model: B-type case}
\begin{figure}[htbp]
\begin{center}
\includegraphics[width=4.4cm]{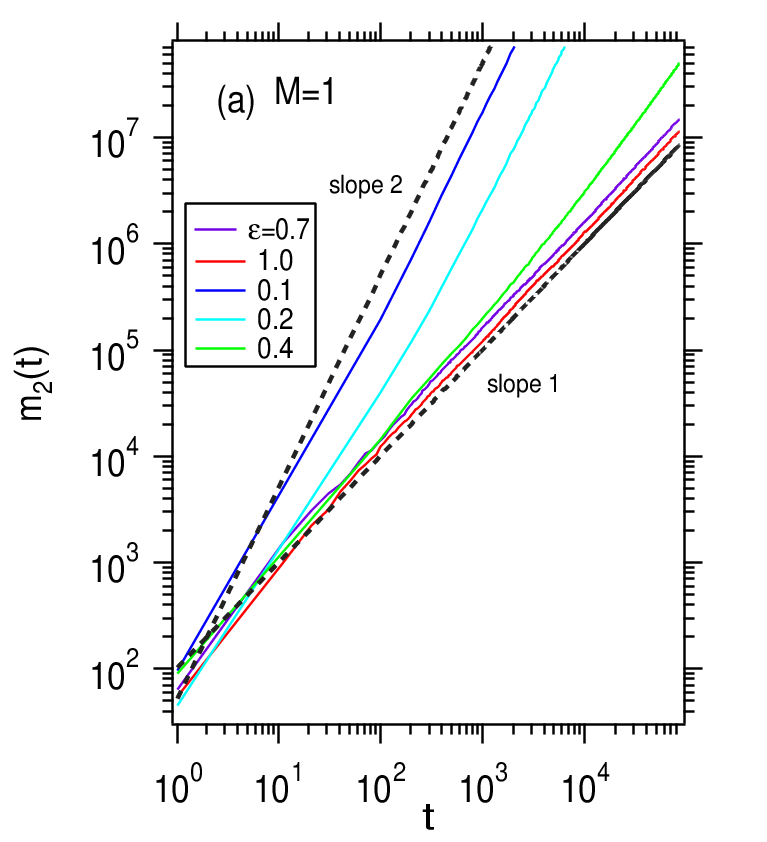}
\hspace{-5mm}
\includegraphics[width=4.4cm]{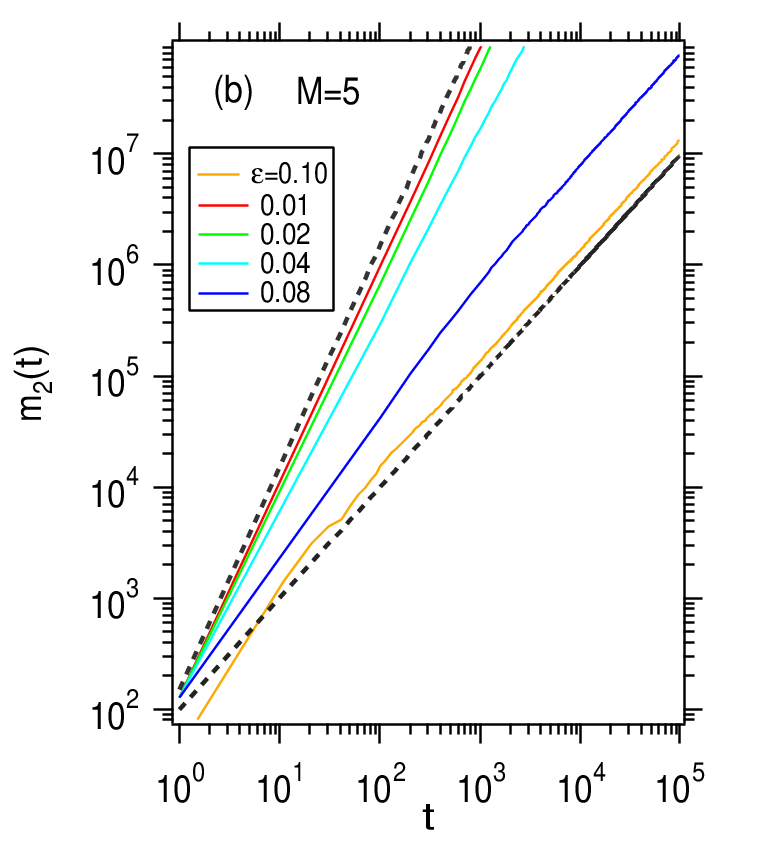}
\caption{(Color online) \label{fig:B-map-c1c5}
The double-logarithmic plots of $m_2(t)$ as a function of $t$ 
for various strength $\eps$ in the  B-type perturbed KHM of $V=1$ 
with (a)$M=1$ and  (b)$M=5$. $\hbar=1/8$.
The values of $\eps$ used are $\eps=0.1, 0.2, 0.4, 0.7, 1.0$ from top to bottom in the panel (a),
and  $\eps=0.01, 0.02, 0.04, 0.08, 0.10$ from top to bottom in the panel (b), respectively.
The dashed lines indicate normal diffusion $m_2 \propto t^1$ 
and ballistic spreading $m_2 \propto t^2$
}
\end{center}
\end{figure}

In the case of the  B-type KHM with $M=1$ and $M=5$, 
the time-dependence of MSD is given in Fig.\ref{fig:B-map-c1c5}.
In both cases, when $\eps$ is small, the growth of $m_2 (t)$ slightly deviates from ballistic spreading 
at the initial time, but it goes to ballistic as $t \to \infty$. 
It can be seen that when $\eps$ becomes large, it gradually approaches 
normal diffusion, $m_2 (t) \sim t^1$, from the initial stage. 
That is, even if $M \geq1$, there is a transition from ballistic to normal diffusion, 
Unlike the LDT case in the previous section, 
there is no theoretical guide based on Anderson transition, 
so the superdiffusion at the BDT must be captured numerically.
This will be shown at another time \cite{yamada22a}.
The asymptotic diffusive behavior is expected as a generic feature 
of decoherence, which takes place when noise is introduced into the system.

\begin{figure}[htbp]
\begin{center}
\includegraphics[width=4.4cm]{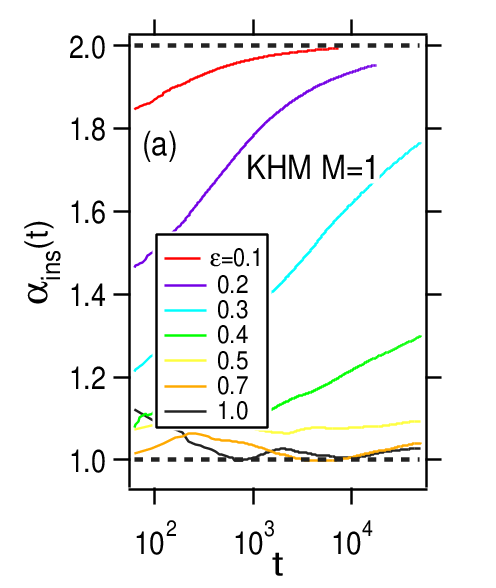}
\hspace{-5mm}
\includegraphics[width=4.4cm]{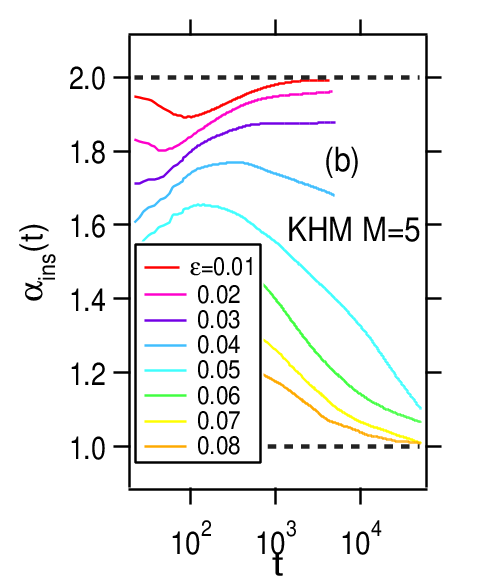}
\caption{(Color online) \label{fig:B-KHM-c1c5}
The time-dependence of $\alpha_{ins}(t)$ 
for various strength $\eps$ in the  B-type perturbed KHM of $V=1$ 
with (a)$M=1$ and  (b)$M=5$. $\hbar=1/8$.
The values of $\eps$ used are $\eps=0.1, 0.2, 0.3, 0.4, 0.5, 0.7, 0.9, 1.0$ 
from top to bottom in the panel (a),
and  $\eps=0.01, 0.02, 0.03, 0.04, 0.05, 0.06, 0.07, 0.08$ 
from top to bottom in the panel (b), respectively.
The dotted lines indicate $\alpha=1$ and $\alpha=2$.
}
\end{center}
\end{figure}

We confirme this by the behavior of $\alpha_{ins}(t)$ used in the LDT in the previous section.
As seen in Fig.\ref{fig:B-KHM-c1c5}, both for M=1 and M=5.
 A transition from $\alpha \to 2$ to $\alpha \to 1$ is suggested 
 when $\eps$ increases, 
  although it is also difficult to accurately determine the value of $\eps_b$ for the transition.


\subsection{Harper model: B-type case}
\begin{figure}[htbp]
\begin{center}
\includegraphics[width=4.4cm]{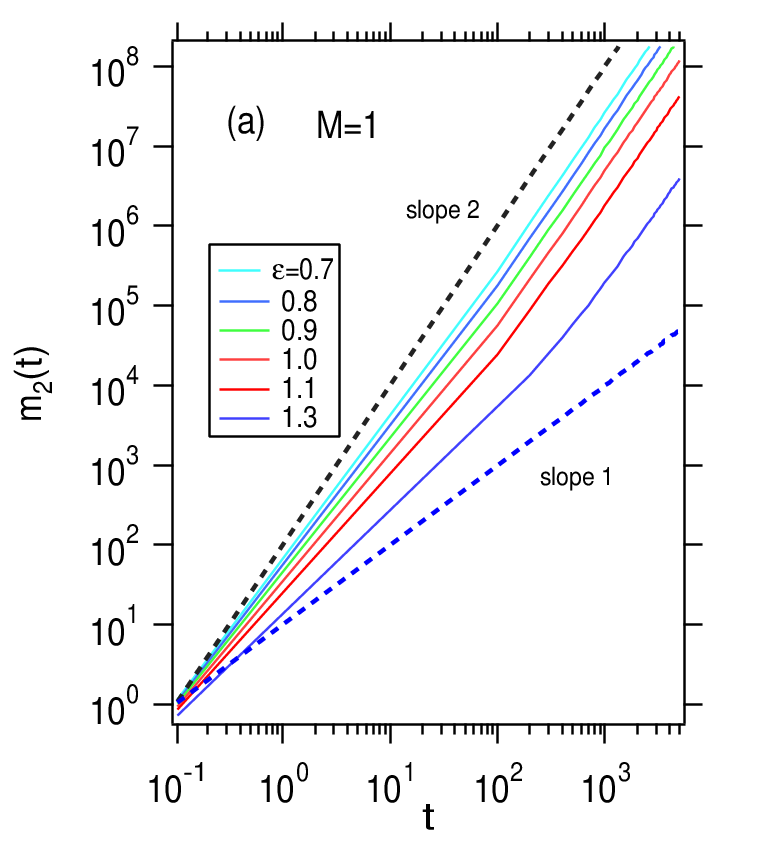}
\hspace{-5mm}
\includegraphics[width=4.4cm]{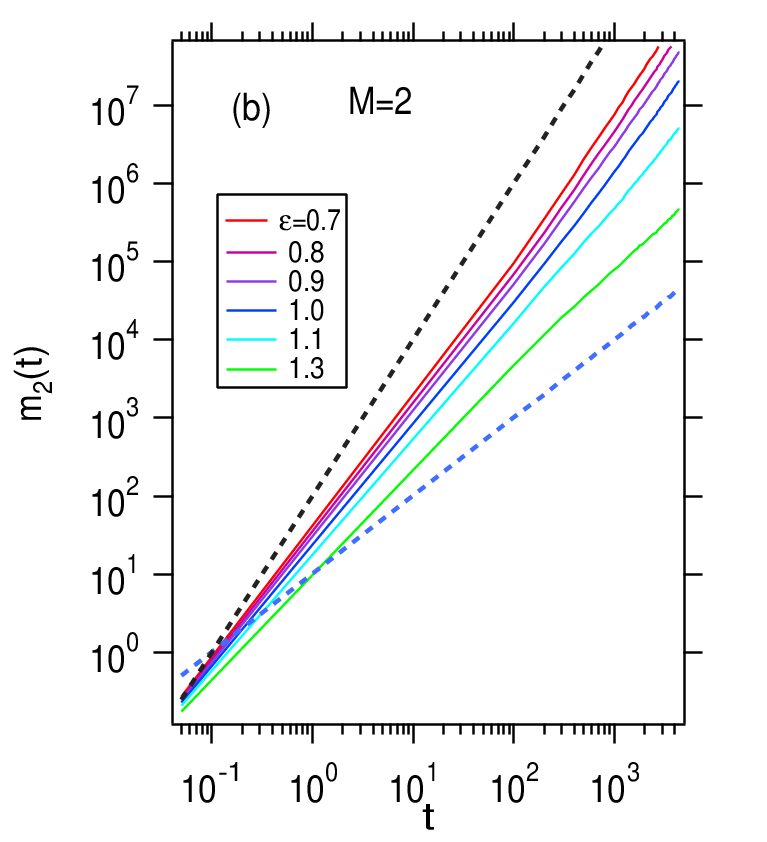}
\hspace{5mm}
\includegraphics[width=4.4cm]{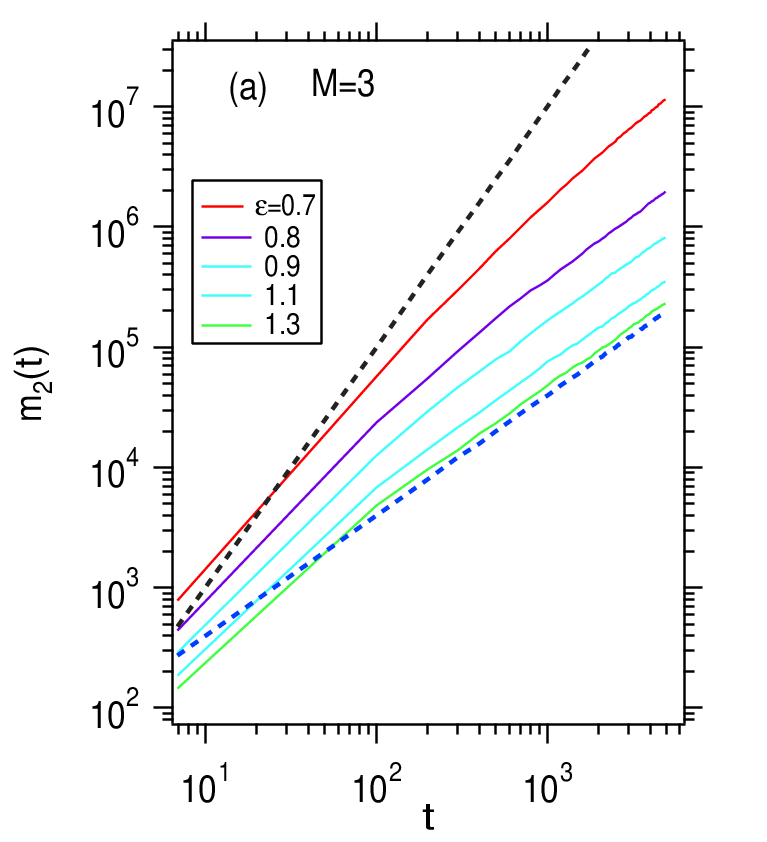}
\hspace{-5mm}
\includegraphics[width=4.4cm]{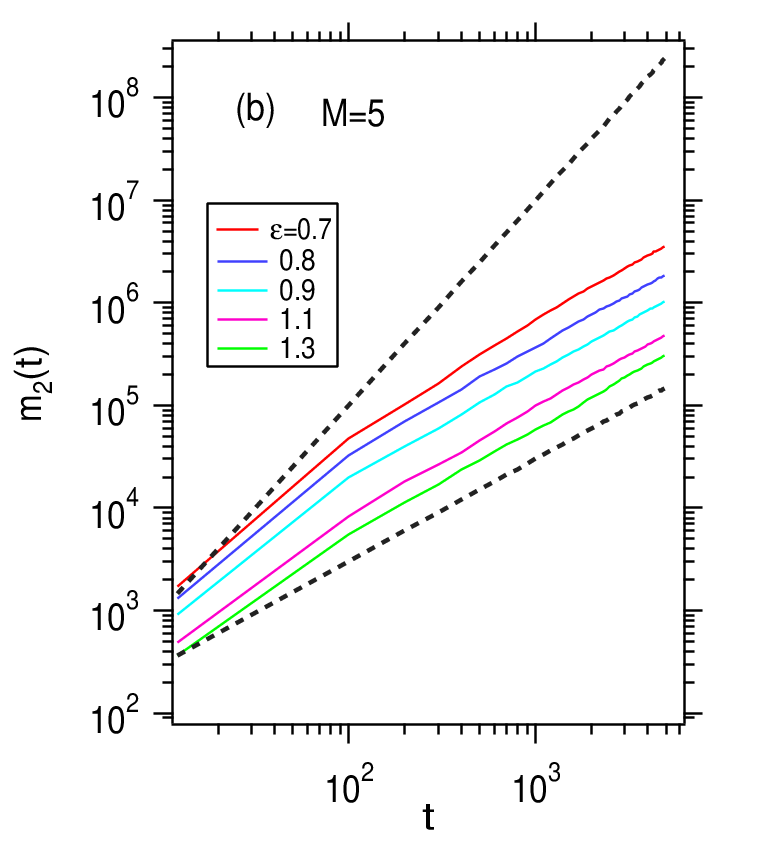}
\caption{(Color online) \label{fig:B-tcont-c1c2}
The double-logarithmic plots of $m_2(t)$ as a function of $t$ 
for various strength $\eps=0.7,0.8,0.9,1.1,1.3$ from top to bottom, 
 in the  B-type perturbed Harper model of $V=1$ 
with (a)$M=1$, (b)$M=2$, (c)$M=3$ and  (d)$M=5$. $\hbar=1/8$.
The dashed lines indicate normal diffusion $m_2 \sim t^1$ 
and ballistic spreading $m_2 \sim t^2$.
}
\end{center}
\end{figure}

\begin{figure}[htbp]
\begin{center}
\includegraphics[width=4.4cm]{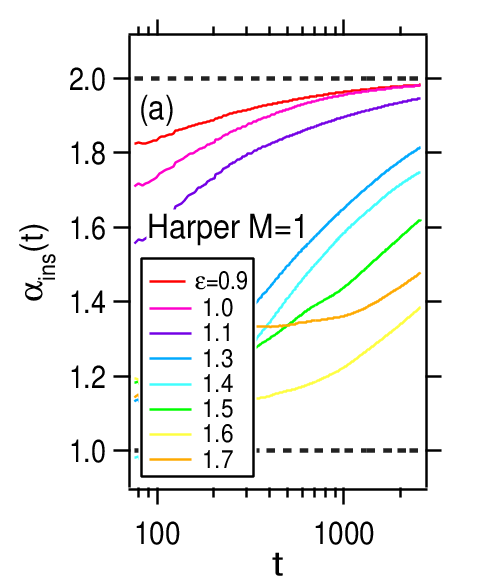}
\hspace{-5mm}
\includegraphics[width=4.4cm]{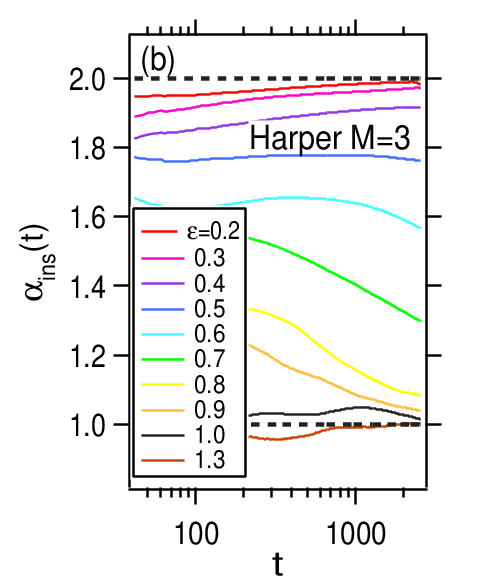}
\caption{(Color online) \label{fig:B-Haper-c1c3}
The time-dependence of $\alpha_{ins}(t)$ 
for various strength $\eps$ in the  B-type perturbed Harper model of $V=1$ 
with (a)$M=1$ and  (b)$M=3$. $\hbar=1/8$.
The values of $\eps$ used are $\eps=0.9, 1.0, 1.1, 1.3, 1.4, 1.5, 1.6, 1.7$ 
from top to bottom in the panel (a),
and  $\eps=0.2, 0.3, 0.4, 0.5, 0.6, 0.7, 0.8, 1.0, 1.3$ 
from top to bottom in the panel (b), respectively.
The dotted lines indicate $\alpha=1$ and $\alpha=2$.
}
\end{center}
\end{figure}

In the case of the  B-type Harper model, 
the time-dependence of MSD for some $\eps$ is given in Fig.\ref{fig:B-tcont-c1c2}.
When $M=1$ in the Fig.\ref{fig:B-tcont-c1c2}(a), it shifts to the normal diffusion side 
at the initial time, but in the all cases, 
we can see that the increase in $m_2(t)$
is parallel to the line with slope 2  as $t \to \infty$. 
On the other hand, as shown in Fig.\ref{fig:B-tcont-c1c2}(b)-(d) for $M\geq2$, 
at short time scale, the wave-packet spread is ballistic, 
but for larger time scale the spread becomes diffusive.
This is a typical feature of approaching the BDT around the critical point $\eps_b$.
The system transits  from ballistic into asymptotic  normal diffusive regime.
Again, let us confirm the features by using  $\alpha_{ins}(t)$.
We can observe the transition from $\alpha \to 2$ to $\alpha \to 1$ for the case of $M=3$
in Fig.\ref{fig:B-Haper-c1c3}.


\section{The delocalized states in A-type and  B-type  systems}
\label{sect:typeAB}
\begin{figure}[htbp]
\begin{center}
\includegraphics[width=6.5cm]{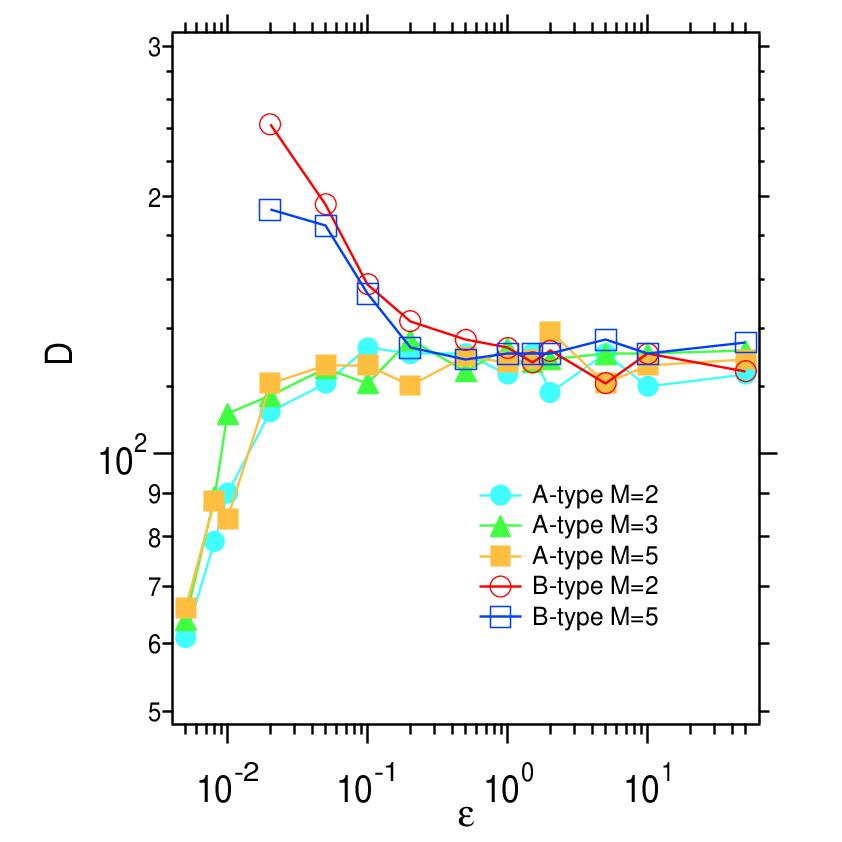}
\caption{\label{fig:Map-AB-Deps.eps}(Color online)
Diffusion coefficient $D$ as a function of $\eps$ in the  A-type and  B-type perturbed 
KHM of $V=5$.$\hbar=1/8$.
Note that the both axes are in logarithmic scale.
}
\end{center}
\end{figure}

It is easy to imagine that the  A-type system asymptotically approaches 
the  B-type system with the increase of $\eps$, because 
the time-dependent part of the Hamiltonian (\ref{eq:Hamiltonian}) can be rewritten:
\beq
\eps V(n)\left[ \frac{L}{\eps}+f(t) \right] \delta_1(t).
\eeq
As seen in the last section, even in the case of  B-type  system, 
the time-dependence of MSD shows the normal diffusion 
for $\eps>\eps_b$  and the diffusion coefficient decreases by increasing $\eps$.
On the other hand, as seen in the previous section, even in  the A-type  system, for $\eps>\eps_c$,
$m_2(t)$ gradually approaches the normal diffusion for  $t \to \infty$.
In this section, we give the relationship 
 between  the results of the diffusive region in the A-type and  B-type systems. 
Here, in the KHM the $\eps-$dependence of the diffusion coefficient in the normal diffusive regime
 in a wide region of $\eps$($>\max\{\eps_c, \eps_b\}$)  is observed in Fig.\ref{fig:Map-AB-Deps.eps}.

In the  A-type system, the diffusion coefficient increases with an increase of $\eps$ 
when $\eps>\eps_c$ and the effect saturates at a certain level. 
On the other hand, in  B-type system, the diffusion coefficient decreases monotonically 
as $\eps$ increases, and it can be seen that $D$ falls to the same level as  A-type system 
for $\eps>>1$. 
For $\eps>0.3$, $M-$dependence of the  diffusion coefficient  has almost disappeared 
for the both systems. 

\begin{figure}[htbp]
\begin{center}
\includegraphics[width=6.5cm]{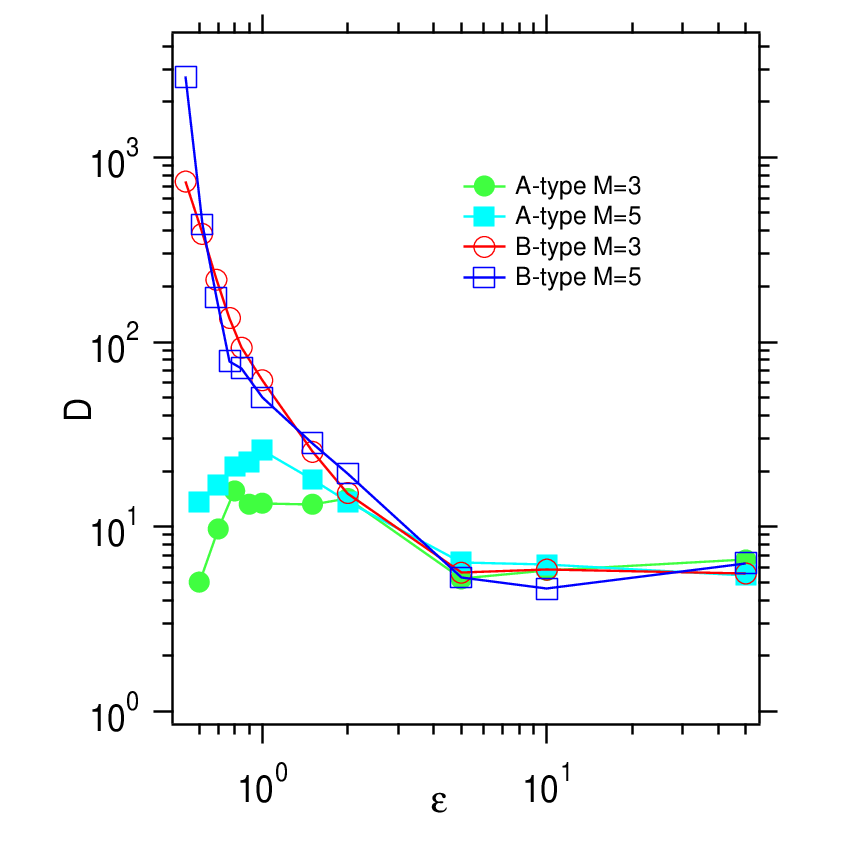}
\caption{\label{fig:tcont-AB-Deps.eps}(Color online)
Diffusion coefficient $D$ as a function of $\eps$ in the  A-type and  B-type perturbed 
 Harper model of $V=1.3$.$\hbar=1/8$.
 Note that the both axes are in logarithmic scale.
}
\end{center}
\end{figure}

Furthermore, as shown in Fig.\ref{fig:tcont-AB-Deps.eps}, similar results are obtained 
for the  A-type and  B-type systems of the time-continuous  Harper model. 
In this case, in  A-type system, increasing $\eps$ also increases $D$ and peaks around $\eps\simeq1$
then decreases, and approaches the same constant level as  B-type system 
for large  $\eps$.

\section{Summary and discussion}
\label{sec:summary}
In the present paper, we investigated the dynamical property 
 of the initially localized wave packet 
 in coherently perturbed kicked Harper and Harper models.

In the  A-type systems ($L=1$), localization-delocalization transition (LDT) 
appeared with increasing the perturbation strength $\eps$. 
The critical value $M_c$ which appear the LDT is $M_c=2$ for KHM, and  
 $M_c=3$ for Harper model.
The property for the number of  the color $M$ is summarized 
in  table \ref{fig:table1}.
The table also describes the LDT 
in the multidimensional Anderson model that corresponds to 
the 1D Anderson map and Anderson model with the quasiperiodic perturbation. 
In the case of KHM, if $M+1$ can be identified with the spatial dimension $d$, the 
 existence of the LDT is a qualitatively consistent result with those of 
the LDT in  $d-$dimensional Anderson model 
\cite{anderson58,abrahams79,lifshiz88,abrahams10}.
However, in the time-continuous systems, 
the critical number of the colors, degrees 
of freedom, is not $d=M+1=2$ but $d=3$, unlike the periodically kicked quantum maps
such as KAM and KHM \cite{notarnicola18,tarquini17}.
We hope that this study will be useful for studying the long-time behavior 
of the dynamics in high-dimensional random systems and multi-degree-of-freedom 
quantum systems that cannot be calculated directly.
Our results suggest that in a general localized systems, there is a transition point $\eps_c$
of the perturbation strength corresponding to the number of color $M$ 
of the quasi-periodic perturbation, and there are characteristic critical dynamics at the point. 
It controls the quantum coherence that causes the localization.

In the  B-type systems ($L=0$) of the KHM and Harper model, the sharp ballistic-diffusive transition (BDT)
 of the dynamics of the wave-packet  can be observed 
with the increase of the strength $\eps$.
The dynamical motion depends on the  number of  the color $M$. The result is  summarized 
in table  \ref{fig:table2}.
In this case, no localized state occurs, but it may be more realistic in terms of
 investigating how the scattering of the one-particle problem changes 
 due to the addition of coherent perturbations to the periodic system of  $\eps=0$. 
Furthermore, it was shown that in both the KHM and Harper model, the  A-type system 
becomes one equivalent to the  B-type system 
 in the normal diffusion region of $\eps >>1$. 
 The asymptotic diffusive behavior is expected as a generic feature 
of decoherence, which takes place when noise is introduced into a system.


This work may lead to a deeper understanding of dynamical localization
and quantum diffusion in quasi-periodic systems.
It also provides insight into 
the control of localized and delocalized states by coherent perturbations in the Floquet engineering.

\begin{table}[htbp]
\begin{center}
 \caption{\label{fig:table1}
 $M-$dependence of the DLT in the  A-type KHM and Harper model.
For $4 \leq M <\infty$ the result is same as the case of $M=3$.
Those in the kicked Anderson and Anderson model are also entered for reference.
The lower lines is a result of the $d-$dimensional disordered systems.
Loc: exponential localization, Diff:Normal diffusion.
}
 \begin{tabular}{lccccc}
\hline
\hline
$M$ & 0 & 1 & 2 & 3 & 4 \\ \hline 
Anderson model \cite{yamada21} & Loc & Loc & Loc & LDT & LDT \\ 
Harper model($V>1$)  & Loc & Loc & Loc & LDT & LDT \\ 
Kicked Anderson model \cite{yamada20}  & Loc & Loc & LDT & LDT  &LDT \\
Kicked rotor \cite{yamada20}  & Loc  & Loc & LDT & LDT  &LDT\\ 
Kicked Harper model ($V>>1$)  & Loc & Loc & LDT & LDT  &LDT \\
 \hline
 d &  d=1 & d=2 & d=3 & d=4 & d=5 \\ \hline 
$d-$D Anderson model    & Loc & Loc & LDT & LDT & LDT\\ \hline
 \end{tabular}
\end{center}
 \end{table} 

\begin{table}[htbp]
\begin{center}
 \caption{\label{fig:table2}
$M-$dependence of the BDT in the  B-type KHM and Harper model.
For $3 \leq M <\infty$ the result is same as the case of $M=2$.
Balli:Ballistic propagation, 
Loc: exponential localization, Diff:Normal diffusion.
}
 \begin{tabular}{lccccc}
\hline
\hline
$M$ &0 & 1 & 2 & 3 & 4 \\ \hline 
Anderson model \cite{yamada21} & Balli & Loc & Diff & Diff & Diff \\
Harper model  ($V>1$) & Balli & Balli & BDT & BDT & BDT \\
Kicked Anderson model \cite{yamada20a} & Balli & Loc & Diff & Diff & Diff \\
Kicked Harper model ($V>>1$)  & Balli & BDT & BDT & BDT & BDT \\ 
 \hline
 \end{tabular}
\end{center}
 \end{table} 


\section*{Acknowledgments}
This work is partly supported by Japanese people's tax via JPSJ KAKENHI 15H03701,
 and the authors would like to acknowledge them.
They are also very grateful to Dr. T.Tsuji and  Koike memorial
house for using the facilities during this study.
The author (H.Y.) would like to acknowlege the hospilatity of the Physics Division of the 
Nippon Dental University at Niigata, where part of this work was completed.

\appendix

\section{Maryland transform}
\label{app:maryland}
We can regard the time-dependent harmonic perturbation $f_\eps(t)$ as the dynamical degrees of freedom. To show this
we introduce the classically canonical action-angle operators 
$(\hatJ_j=-i\hbar \frac{\pr_j}{\pr_j\phi_j}, \phi_j)$
representing the harmonic perturbation as the linear modes, and we 
 call them the color modes.
 Each quantum oscillator has the action eigenstates $|n_j>$ 
 with the action eigenvalue $J_j=n_i\hbar~(n_j:$integer) 
and the energy $n_j\hbar\omega_j$, 
where $\hatJ_j|m_j\>=m_j\hbar|m_j\>$($m_j\in {\Bbb Z}$).
Thus the system (\ref{eq:Hamiltonian}) is 
regarded as a quantum  system of $(M+1)$-degrees of freedom 
spanned by the quantum states $|n>\prod_{j=1}^M |n_j>$. 
Then  the Hamiltonian $\tilde{H}_{kick}$ that include the color modes becomes
\beq
&& \tilde{H}_{kick}(\hatp,\hatq,\{\hatJ_j\},\{\hatphi_j\}) =2\cos(\hatp/\hbar)+ \nn \\
&& 
2V\cos(2\pi Q \hatq) \left[L+ \frac{\eps}{\sqrt{M}} \sum_j^M \cos \hatphi_j \right]\delta_1(t)
+\sum_{j=1}^M \omega_j\hatJ_j.
\eeq

Let us consider an eigenvalue equation 
\beq
  \e^{-i\hatA}\e^{-i\hatB}\e^{-i\hatC}|u\> =\e^{-i\gamma}|u\>, 
\label{eq:eigen-value-problem}
\eeq
where 
\beq
\begin{cases}
\hatA =(2VL\cos(2\pi Q\hatq)+\sum_j^M\omega_i \hatJ_j)/\hbar , \\
\hatB =\cos(2\pi Q\hatq)\frac{\eps 2V}{\sqrt{M}}\sum_j^M\cos\hatphi_j/\hbar , \\
\hatC =2\cos (\hatp/\hbar)/\hbar 
\end{cases}
\eeq
for the time-evolution operator
$T\e^{-\frac{i}{\hbar}\int \tilde{H}_{kick} dt}$.
$\gamma$ and $|u\>$ are the quasi-eigenvalue and quasi-eigenstate.
Here, if the eigenstate representation of $\hatJ_j$ is used,
we can obtain the following 
$(M+1)-$dimensional tight-binding expression 
by the Maryland transform \cite{yamada20}:
\beq
\label{eq:Maryland_AM}
& & D(n,\{m_j\})u(n,\{m_j\}) +  \nn \\
& & \sum_{n',\{m_j^{'}\}}\<n,\{m_j\}|\hat{t}_{KHM}|n',\{m_j^{'}\}\>  
u(n',\{m_j^{'}\}) =0, 
\eeq
where $\{m_j\}=(m_1,....,m_M)$.
Here the diagonal term is 
\beq
D(n,\{m_j\})=
\tan \left[ \frac{2VL\cos(2\pi Q n)+\hbar\sum_j^Mm_j\omega_j}{2\hbar}-\frac{\gamma}{2}
 \right], 
\label{eq:diagonal}
\eeq
and the $\hat{t}_{KHM}$ of the off-diagonal term is 
\beq
\hat{t}_{KHM}=i\frac{
e^{-i  \frac{\eps 2V}{\sqrt{M}}\cos(2\pi Q\hatq) (\sum_{j}^{M}\cos\hatphi_j)/\hbar}-
e^{i2\cos(\hatp/\hbar)/\hbar}
}
{e^{-i\frac{\eps 2V}{\sqrt{M}}\cos(2\pi Q\hatq) (\sum_{j}^{M}\cos\hatphi_j) /\hbar}+
e^{i2\cos(\hatp/\hbar)/\hbar}
}.
\eeq
$D(n,\{m_j\})$ is $\eps-$independent and the off-diagonal term is $\eps-$dependent.
It follows that the $(M+1)-$dimensional tight-binding model
of the KHM have singularity of the on-site energy caused by 
tangent function and long-range hopping caused by the kick $\delta_1(t)$.


In the cases of  the A-type system  ($L=1$), 
there exists $2V^*=2\pi\hbar$, that is $V^* \simeq 0.38$ for $\hbar=1/8$, where the effect of the
the fluctuation width of the diagonal term is saturated
for the change of  the potential strength $V$ due to the tangent function (\ref{eq:diagonal}).
Therefore, for $V> V^*$
the delocalization can be caused by  increase of $\eps$ in the off-diagonal term in the form of $2\eps V$.
Note that  the transform is not 
possible in the time-continuous systems such as Harper model.





\begin{thebibliography}{00}
\bibitem{math-comment}
However, there is no mathematical proof of the LDT for the wavepacket dynamics 
except for the localization case.
In addition, it is proven that in the one-dimensional tight-binding model with the potential with a finite period
the spectrum is absolutely continuous and wavepacket shows ballistic motion
$m_2 \sim t^2$. 
\end{thebibliography}
\end{document}